\documentclass[secnumarabic, twocolumn, graphics,floatfix, superscriptaddress,tightenlines, aps, prl]{revtex4-1}
\usepackage[dvips]{graphicx}
\usepackage{dcolumn}
\usepackage{bm}
\usepackage{color}

\usepackage[dvips]{graphicx}
\usepackage{amssymb,amsmath}

\graphicspath{{figs/}}

\begin{document}

\title{Influence of magnetic quantum confined Stark effect on the spin lifetime of indirect excitons.}
\author{P.~Andreakou}
\affiliation{Laboratoire Charles Coulomb, UMR 5221 CNRS/ Universit\'{e} Montpellier 2,
F-34095, Montpellier, France}
\author{A.~V.~Mikhailov}
\affiliation{Laboratoire Charles Coulomb, UMR 5221 CNRS/ Universit\'{e} Montpellier 2,
F-34095, Montpellier, France}
\affiliation{Spin Optics Laboratory, St-Petersburg State
University, 1, Ulianovskaya, St-Peterbsurg, 198504, Russia}
\author{S.~Cronenberger}
\affiliation{Laboratoire Charles Coulomb, UMR 5221 CNRS/ Universit\'{e} Montpellier 2,
F-34095, Montpellier, France}
\author{D.~Scalbert}
\affiliation{Laboratoire Charles Coulomb, UMR 5221 CNRS/ Universit\'{e} Montpellier 2,
F-34095, Montpellier, France}
\author{A.~Nalitov}
\affiliation{School of Physics and
Astronomy, University of Southampton, Southampton, SO17 1BJ, United
Kingdom}
\author{A.~V.~Kavokin}
\affiliation{Spin Optics Laboratory, St-Petersburg State
University, 1, Ulianovskaya, St-Peterbsurg, 198504, Russia}
\affiliation{School of Physics and
Astronomy, University of Southampton, Southampton, SO17 1BJ, United
Kingdom}
\affiliation{Russian Quantum Center, 100, Novaya, Skolkovo, Moscow
Region, 143025, Russia}
\author{M.~Nawrocki}
\affiliation{Institute of Experimental Physics, University of
Warsaw, $Ho\dot{z}a$ 69, 00-681 Warsaw, Poland}
\author{L.~V.~Butov}
\affiliation{Department of Physics, University of California at San Diego, La Jolla, CA
92093-0319, USA}
\author{K.~L.~Campman}
\affiliation{Materials Department, University of California at Santa Barbara, Santa
Barbara, California 93106-5050, USA}
\author{A.~C.~Gossard}
\affiliation{Materials Department, University of California at Santa Barbara, Santa
Barbara, California 93106-5050, USA}
\author{M.~Vladimirova}
\affiliation{Laboratoire Charles Coulomb, UMR 5221 CNRS/ Universit\'{e} Montpellier 2,
F-34095, Montpellier, France}

\begin{abstract}
We report on the unusual and counter-intuitive behaviour of spin lifetime of excitons in coupled semiconductor quantum wells (CQWs) in the presence of in-plane magnetic field. Instead of conventional acceleration of spin relaxation due to the Larmor precession of electron and hole spins we observe a strong increase of the spin relaxation time at low magnetic fields followed by saturation and decrease  at higher fields.
We argue that this non-monotonic spin relaxation dynamics is a fingerprint of the magnetic quantum confined Stark effect.
In the presence of electric field along the CQW growth axis,  an applied magnetic field efficiently suppresses the exciton spin coherence, due to inhomogeneous broadening of the $g$-factor distribution.

\end{abstract}

\pacs{} \maketitle

\section{Introduction}

The effect of magnetic field on Wannier-Mott excitons is studied since late 1950s. Theoretical works by Elliott and Loudon \cite{Elliott}, Hasegawa and Hovard \cite{Hasegawa}, Gor'kov and Dzyaloshinskii \cite{Gorkov} describe the diamagnetic energy shift and fine structure of bulk excitons. Lerner and Lozovik expanded these studies to two-dimensional systems including quantum wells (QWs) \cite{LernerLozovik}. Later,  a detailed theory has been developed for magneto-excitons in biased coupled quantum wells (CQWs), where both spatially direct (DX) and indirect (IX) exction states may be realised \cite{Gorbatsevich,Butov2000,Lozovik2002}. It has been shown that due to the joint action of normal-to-QW-plane electric and in-plane magnetic fields the exciton  dispersion can be strongly affected. 
Besides this, magnetic fields strongly affect the exciton oscillator strength, that is ability to absorb or emit light, as has been pointed out in the seminal paper of Thomas and Hopfield \cite{Hopfield}. Due to the opposite orientations of Lorentz forces acting upon electron and hole, the exciton acquires a stationary dipole moment in the presence of a magnetic field. 
This constitutes the magnetic Stark effect studied in various semiconductor systems \cite{Hopfield, Lafrentz2013}.
The conventional Stark effect is enhanced in CQWs compared to single QWs \cite{Armiento}.
In this paper we address the magnetic Stark effect in CQWs.
\begin{figure}[t!]
\center{\includegraphics[width=0.8\linewidth]{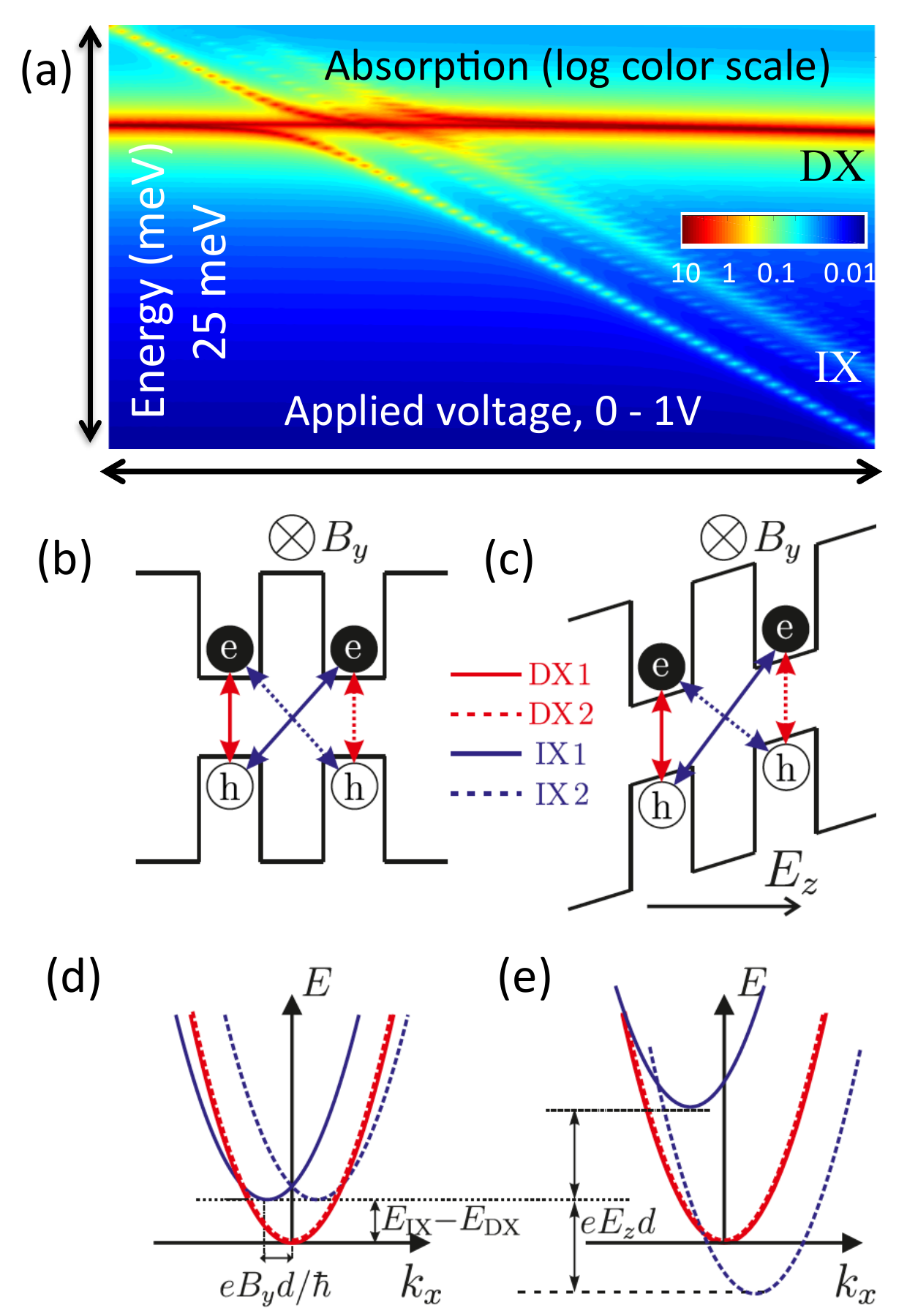} }
\caption{(color online) (a) Colormap of the excitonic absorption in a CQW structure studied in this work, 
calculated as a function of applied electric bias. 
Sketch of the CQW band structure in the presence (c), (e), or absence (b), (d) of electric bias along $z$-axis.
Red and blue parabolas in (d-e) are dispersions of two direct and two indirect states  corresponding to optical transitions indicated in (b-c).
}
 \label{fig1}
\end{figure}
%
%

In CQWs, excitonic states are mixtures of traditional intrawell, or DX states
and interwell, or IX states, consisting of an electron and a hole confined in different QWs, Fig. \ref{fig1}~(b, c).
This mixing can be controlled by the external bias. 
%
%
The corresponding exciton energies and oscillator strengths  can be
accurately calculated by solving Schr\"{o}dinger equations for
different values of the gate voltage \cite{Sivalertporn}.
An example of such calculation for a typical CQW sample is given in Fig.
\ref{fig1}~(a), where a colormap of the excitonic absorption 
in a CQW structure is shown in the energy/gate voltage plane.
The absorption is inversely proportional to the exciton lifetime.
%
%
In the absence of applied bias, DX-like state is the ground state of the system, IX is 
several meV above it, and its oscillator strength is only $10$ times
lower.
%
%
By contrast, at strong gate voltage, the IX state is about $20$~meV below
the DX and has an oscillator strength $100$ times lower than the DX.
At intermediate gate voltages, IX and DX
states anticross.
It is convenient to describe the system in terms of  direct and  indirect states, interacting via electron tunneling.
%

The spin lifetime of a pure DX state is short, $\approx 50$~ps.
Indeed, an important source of DX spin relaxation is the fluctuating
effective magnetic field, which originates from the momentum dependent component 
of the exchange interaction \cite{MaillaleSham}. 
The fluctuations, due to the scattering of the exciton center of mass momentum, 
are responsible for the exciton spin relaxation, in the same manner as any other 
motional narrowing spin-flip processes are, with the characteristic dependence of the 
spin-relaxation time on the inverse momentum scattering time.
For pure IX the exchange interaction  is negligible due to the low electron-hole overlap, so that the fluctuating wave-vector dependent exchange field does not affect spin relaxation.
At low exciton densities, IX spin lifetime can be even longer than the spin relaxation time of a two-dimensional electron gas (2DEG) 
systematically present in biased QWs \cite{Andreakou}.
This difference is essentially due to stronger localisation of 
IXs in the QW disorder potential, as compared to 2DEG or DX. 
The variation of the spin lifetime of excitons with the gate voltage 
can be understood as being due to the mixing between the purely DX state characterized by a fast 
relaxation rate, and the purely IX state having a slow spin relaxation rate \cite{Andreakou}.
%
%
%
%

Application of a magnetic field in the plane of the CQWs provides a rich playground 
where the combination of magnetic and traditional Stark effect, disorder, interactions and mobility governs the spin dynamics in the system.
Indeed, in-plane magnetic field shifts the dispersion of IX states in $k$-space, 
as illustrated  schematically 
in Figure \ref{fig1}~(d, e) \cite{Butov2000, Gorbatsevich}.
%
%
In this work we study the implication of this phenomenon for the spin dynamics in CQWs, 
using the time-resolved Kerr rotation spectroscopy.
%
In the presence of an in-plane magnetic field this technique allows us to determine the transverse spin lifetime, which is limited by the exciton recombination time, the spin coherence time, and the eventual pure spin dephasing due
to an inhomogeneous broadened distribution of the g-factors. 
We identify two regimes of spin coherence, controlled by the strength of the applied electric field. 
At strong bias, zero-field spin lifetime reaches $10$~ns.
The applied magnetic field leads to exciton spin dephasing, 
due to strong inhomogeneous broadening of the $g$-factor distribution in biased CQWs.
A similar behaviour is observed in the 2DEG in the hoping regime \cite{Zhukov2006, Zhukov2007}. 
At zero bias, application of the in-plane magnetic field results in a strong increase of exciton spin  lifetime up to $5$~T, followed by a decrease at higher fields and a non-monotonic behaviour of the spin relaxation time.
We interpret this unusual behaviour  as a consequence of the magnetic Stark effect, which in CQWs
converts DX to IX, having much slower spin relaxation rate, while the distribution of g-factors plays much weaker role at zero bias.

\section{Sample and experimental setup}
\label{setup}
Our sample consists of two $8$ nm wide GaAs quantum wells separated
by a $4$ nm Al$_{0.33}$Ga$_{0.67}$As barrier and surrounded by $200$ nm Al$%
_{0.33}$Ga$_{0.67}$As layers.
The voltage $V_{g}$ applied between the conducting n-GaAs layers
drops in the insulating layer between them \cite{Butov1999}.
The sample is placed in the helium bath magneto-optical cryostat.

We perform photoinduced Kerr
rotation and  reflectivity experiments at $2$~K.
Two-color measurements are realized by spectral filtering of pump
and probe pulses. The pulse duration is $1$ ps, the spectral width
is $1.5$ meV. The Ti-Sapphire laser repetition rate is reduced to
$20$~MHz in order to avoid exciton accumulation between pulses at high gate voltage
and high magnetic field.
Typical powers are $120$ and $70$ $\mu $W for pump and probe,
respectively, focused on a $100$~$\mu $m diameter spot \footnote{We
have checked that reducing probe to pump power ratio does not change
the signal dynamics.}. Magnetic fields are applied in the plane of the structure (Voigt geometry).
Spin-polarized DXs are optically excited in the CQW by a circularly
polarized pump pulse, tuned in the vicinity of DX resonance. The resulting dynamics of
the spin polarization (exciton density) is monitored via Kerr rotation (reflectivity)
 of the delayed linearly polarized probe pulse. 
 
 The probe energy is also tuned around the DX resonance, 
 and is chosen independently of the
 pump energy, in order to optimise the signal.
 We have shown in our previous work, that the spin dynamics of  DXs,
IXs, and residual 2DEG in coupled quantum
wells can be efficiently addressed in this configuration \cite{Andreakou}.

\begin{figure}[tbp]
\center{\includegraphics[width=1\linewidth]{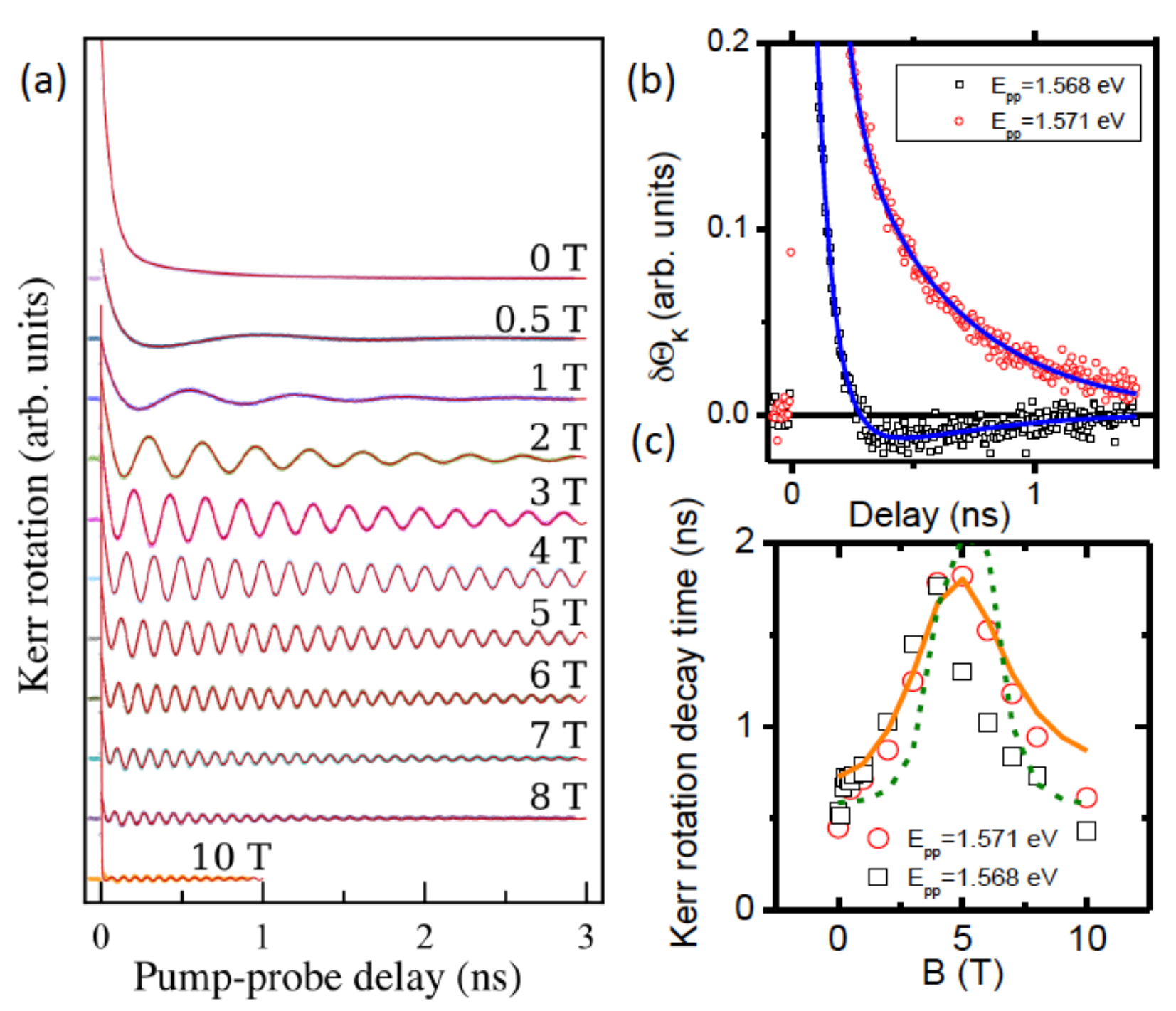} }
\caption{ (a) Waterfall plot of Kerr rotation measured at zero electric bias as a function of the pump-probe delay  at different magnetic field intensities,  $E_{pp}=1.571$~eV,  $E_{pr}=1.569$~eV. (b) Same measurements at $B=0$ for two different pump energies. (c) Magnetic field dependence of the slowest relaxation time measured at zero electric bias for two different excitation energies. Lines are fit to two models, based on the Liouville  equation with Lindblad term (solid line) and on the microscopic analysis of the Schr\"{o}dinger equation (dashed line). } \label{figVg0}
\end{figure}

\section{results}

Figure \ref{figVg0} presents the main result of this work. It shows the Kerr rotation signal measured at the pump energy $E_{pp}=1.571$~eV and the probe energy  $E_{pr}=1.569$~eV for different values of magnetic field ranging from zero to $10$ T. No gate voltage is applied, pump and probe energies were chosen to optimise the zero-field signal, and kept fixed for the set of measurements shown in Fig. \ref{figVg0}~(a). 
One can see that the monotonous bi-exponential decay at $B=0$ is replaced by a much longer living oscillatory behaviour once the magnetic field increases up to about $5$T. The further increase of the magnetic field is accompanied by the decrease of the decay time, back to the zero-field value. All the curves measured in the presence of the magnetic field are well described by a fast ($50$~ps) exponential decay, followed by a slower decaying cosine function.

We have shown in our previous work that at low excitation energy and power, one can reach the regime where 
exciton spin precesses even in the absence of the applied magnetic field \cite{Andreakou}. 
This precession is due to a small splitting between
two perpendicularly polarized linear exciton states $\delta _{xy}$
that is generally present in QW structures \cite{KrizhanovskiiPRB2006,HighPRL2013}.
For exciton spin this splitting acts as an effective in-plane
magnetic field. Therefore, relaxation of the spin polarization is
accompanied by its rotation around this effective field.
Such precession is very sensitive to the excitation energy and is only observed at low pump energy and power, as shown in Fig. \ref{figVg0}~(b). 
At low pump energy, when excitons are essentially localised, we observe a decay accompanied by an oscillation of the Kerr rotation signal, while at high energy pumping simple bi-exponential decay is observed. 
The set of measurements shown in Fig. \ref{figVg0}~(a) corresponds to the high energy excitation, where there is no spin precession at $B=0$.
The corresponding decay times of the precessing component extracted from the fitting procedure are shown in Fig. \ref{figVg0}~(c).
The non-monotonous behaviour as a function of the applied magnetic field can be clearly observed.  
It is robust with respect to the 
excitation energy, as shown in Fig. \ref{figVg0}~(c), and persists whatever the pump power and energy is.

We attribute the oscillatory behaviour of the Kerr rotation signal to the precession of the electron spin, rather than to the exciton spin precession.
Indeed, in GaAs-based QWs such precession has already been observed  \cite{Amand1997}. It was shown,
that when the hole spin relaxation time  $\tau_h$ is shorter than $\hbar/\Delta_0$, the spin of an electron bound into an exciton precesses at the same frequency as the free electron spin
 \cite{Dyakonov1997}. 
Here $\Delta_0$ stands for the short-range part of the exchange interaction. 
In our $8$~nm QWs, $\Delta_0 \approx 70$~$\mu$eV \cite{Blackwood1994},  and, at least for delocalised excitons hole spin relaxation is fast $\tau_h<10$~ps.
Therefore, we 
conclude that the observed spin dynamics should be attributed to the precession of the spin of electrons bound to holes within  excitons.
We have checked, that in the regime where zero-field precession of exciton spin is observed, application 
of the magnetic field of only $0.15$~T is sufficient to overcome the hole exchange field acting on the electron spin, and recover electron spin precession.
Therefore, in what follows we consider that Kerr rotation oscillations observed 
at zero bias are related to the spin precession of electrons bound within excitons.

There are two very surprising findings shown in Fig. \ref{figVg0}. First of all, except for indirect excitons, typical relaxation times observed for  excitons in GaAs QWs do not exceed $100$~ps \cite{DyakonovBook}, while in the present experiment we deal with much
longer times.
 Moreover, the spin lifetime is not expected to increase when magnetic field increases. 
 A constant or decreasing with in-plane magnetic field spin lifetime is typically observed in semiconductor QWs \cite{Kikkawa1998, Zhukov2006}. 
 This is due to dephasing governed by the width of the electron $g$-factor distribution 
 $\tau_{inh}(\Delta g, B)=\sqrt{2}\hbar/(\Delta g \mu_B B)$, and it is inversely proportional to the applied 
 magnetic field, in a striking contrast with our experimental observation. 

\begin{figure}[tbp]
\center{\includegraphics[width=1\linewidth]{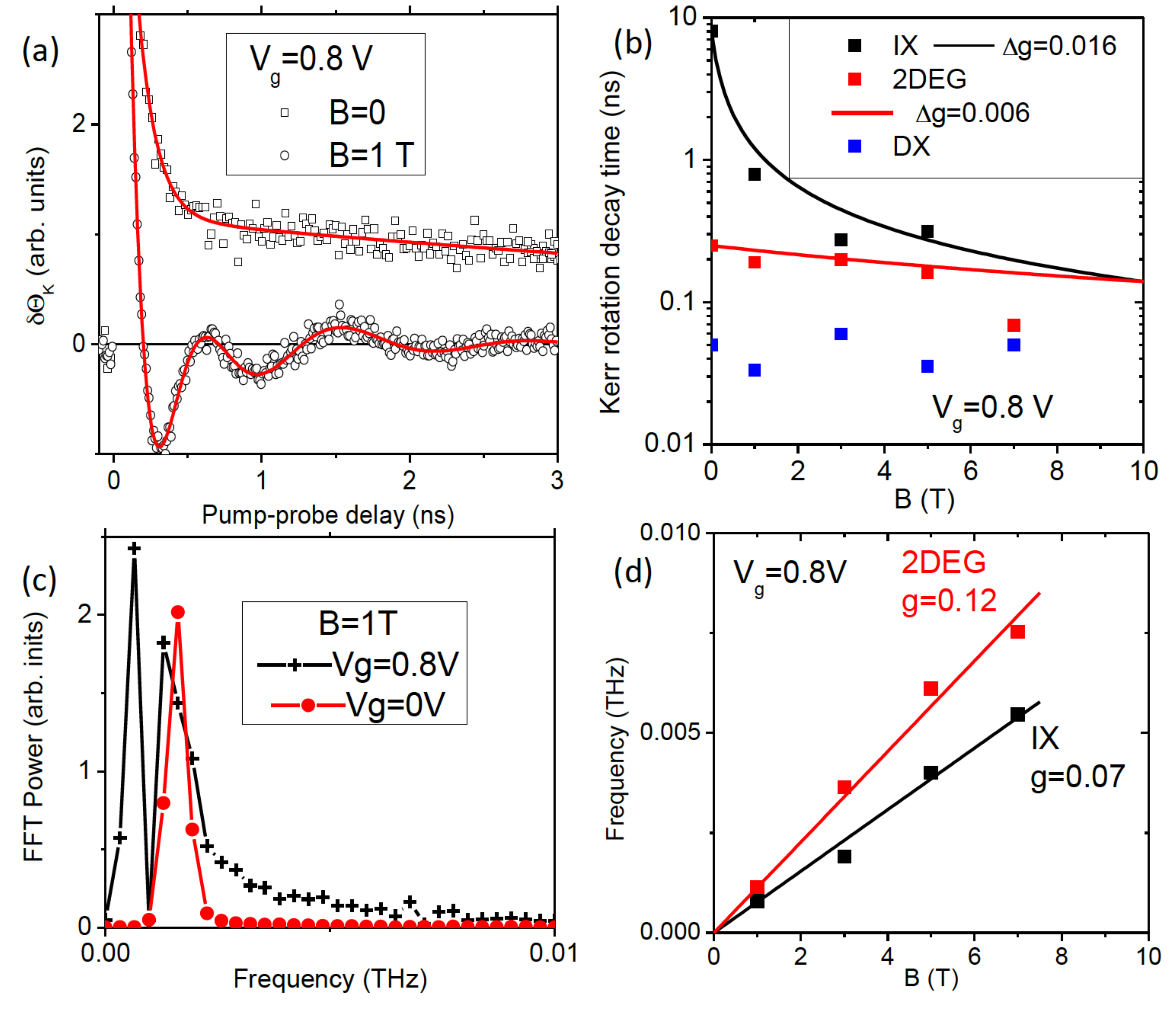} }
\caption{(a) Kerr rotation measured at $V_g=0.8$~V as a function of the pump-probe delay  at $B=0$ and $B=1$~T, 
$E_{pp}=1.568$~eV,  $E_{pr}=1.569$~eV.
(b) Three characteristic decay times extracted from Kerr rotation measurements at $V_g=0.8$~V and at different in-plane magnetic fields. These decay times are ascribed to DX, IX, and 2DEG spin relaxation. 
Solid line are fit the spin dephasing model, assuming $g$-factor distribution $\Delta g=0.016$ for IX and  $\Delta g=0.006$
for 2DEG.
(c) Fourier spectra of Kerr rotation measured at $B=1$~T. Zero bias spectrum is compared to $V_g=0.8$~V.
(d) Two precession frequencies extracted from Kerr rotation measurements at $V_g=0.8$~V and at different in-plane magnetic fields. These frequencies are ascribed to IX, and 2DEG spin rprecession} \label{figVg}
\end{figure}

Before going into the analysis of the $V_g=0$ results, let us now consider the spin coherence in the presence of the strong gate voltage $V_g=0.8$~V, as shown in  Fig. \ref{figVg}. 
In this regime, IX is the lowest energy exciton state of the system and in the Kerr rotation signal at $B=0$ 
we observe three exponentially decaying components (squares in Fig. \ref{figVg}~(a)) \cite{Andreakou}.
They can be attributed to the spin relaxation of the DX, 2DEG,
which forms in biased CQWs \cite{Butov2004,Rapaport2004,ButovJETP},
and IXs. 
The latter has much longer spin relaxation time, up to $10$~ns.
At $B=1$~T we still observe three components (circles in Fig. \ref{figVg}~(a)). The experimental data are fitted to a linear superposition of one exponential and two damped cosine functions. 
The fastest exponential decay, associated with the DX spin, is not affected by magnetic field, as it is much faster than
 the precession period.
The two other components exhibit the oscillatory behaviour with different precession frequencies and decay times.
This is illustrated in \ref{figVg}~(c), were Fourier spectra of Kerr rotation measured at $B=1$~T are shown.
While at zero bias  only one peak appears in the Fourier spectrum  at  $V_g=0.8$~V two peaks can be clearly distinguished,
the lowest frequency corresponding to the slowest decay.
The precession frequency of the slow component is related to the IX spin (more precisely to the precession of the spin of electron bound into IX), and the fast component is associated with the bare electron spin.
The magnetic field dependence of the two precession frequencies is shown in Fig. \ref{figVg}~(d)  \footnote{It was difficult to mesure the decay times at $B>7$~T, due to the smaller intensity of the signal}.
One can see that it corresponds to different g-factors, which may arise from the different mass and density and therefore different localization of IXs and electrons \cite{Andreakou}.
Indeed, the degree of localization is a crucial parameter, that controls the $g$-factor values in 
GaAs/AlAs-based heterostructures \cite{Cundiff}. 
Note, that in undoped CQWs identical to the one studied here, previous studies have found 
the same value of the $g$-factor $g=0.12$ \cite{Poggio2004}.
The decay times obtained at $V_g=0.8$~V  are shown in Fig. \ref{figVg}~(b).
While the shortest (DX) spin lifetime remains constant, both 2DEG and IX spin lifetimes
decrease with the increase of the magnetic field. Solid lines show the fit to $1/B$ behaviour, consistent with the inhomogeneous broadening expected from the distribution of $g$-factors, which gives $\Delta g = 0.006$ for a 2DEG and a higher value, $\Delta g = 0.016$,  for more localised IXs. 

\begin{figure}[tbp]
\center{\includegraphics[width=1\linewidth]{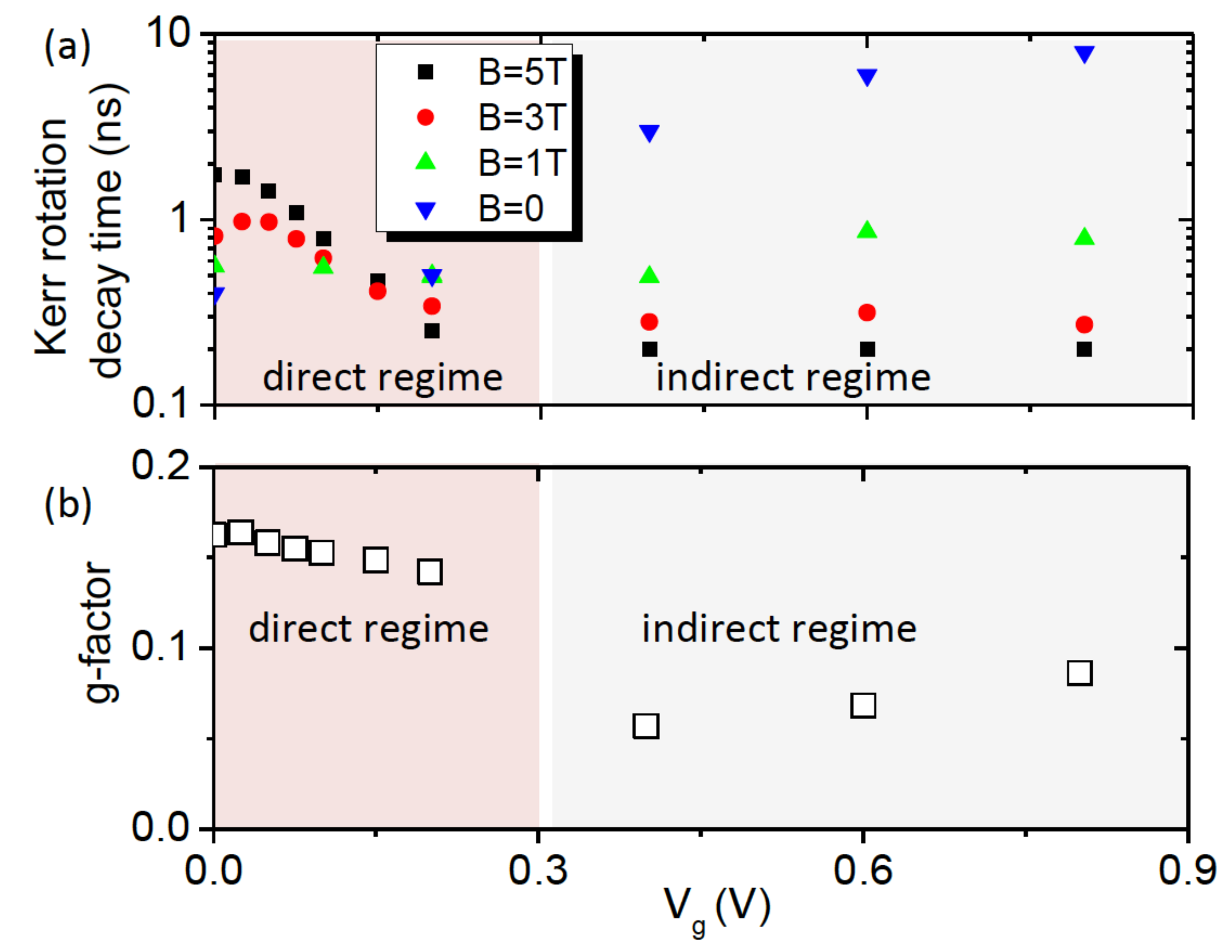} }
\caption{Gate voltage dependence of  (a) the slowest decay time measured in Kerr rotation 
scans at different gate voltages and in-plane magnetic fields (b) Gate voltage dependence of  g-factor, associated with this 
slowest component. 
Two regimes can be identified.
At sufficiently small $V_g$, such that DX state remains below IX (direct regime) , decay time increases with magnetic field.
Under gate voltage such that IX becomes the lowest exciton state in the system (indirect regime), the dependence is inverse,
decay time drops dramatically in the presence of the magnetic field. } \label{figB}
\end{figure}

Let us summarise the dependence of the  spin lifetime in a system of CQWs 
on both the in-plane magnetic field and the electric field along the growth axis (defined by the applied gate voltage). %
Fig. \ref{figB}~(a) shows the longest spin lifetime extracted from the fitting procedure described above. The values are given for different  in-plane magnetic fields up to $5$~T, for each field the gate voltage dependence is shown.
Two regimes can be distinguished. They are indicated by different background colors in  Fig. \ref{figB}.
At low voltage, in the regime where IX energy is higher than that of DX (direct regime), spin relaxation time increases 
with the magnetic field increase. The corresponding $g$-factor slightly increases with bias but 
remains above $g=0.1$, Fig. \ref{figB}~(b).
Using photo-induced reflectivity technique described in Ref. \cite{Andreakou}, we could not detect any measurable modification of the exciton lifetime with magnetic field in this regime, it remains slightly below $10$~ns up to $10$~T. %
At high voltage, i. e. above $\approx 0.3$~V,  IX becomes the lowest energy exciton state (indirect regime) \cite{Butov19992}.
In this regime the $g$-factor decreases substantially and spin lifetime time also changes its behaviour. 
It decreases when magnetic field increases. 
It was shown, that in this regime, magnetic field also leads to strong increase of the IX lifetime, due to the shift of  the IX dispersion in $k$-space \cite{Butov2000}. 
This feature is also reproduced in our photo-induced reflectivity experiments. At $V_g=0.8$~V and $7$~T exciton lifetime approaches the time delay between the laser pulses 
($48$~ns),  leading to the accumulation of excitons in the structure.
The $1/B$ behaviour  of the spin lifetime in the indirect regime can be understood in terms of the inhomogeneous broadening of the 
long-living and strongly localized IX as shown in Fig. \ref{figVg}~(b).
Strong gate voltages pushes electron and hole towards QW interfaces, which contributes to the increasing role of the disorder potential.
Additional localisation also leads to the decreasing $g$-factor \cite{Cundiff}.
By contrast, the increase of the spin lifetime in the presence of the in-plane magnetic field observed in the direct regime does not have analogs in other electronic or excitonic systems.
We will show in the theoretical part of the paper, that this effect is due to the magnetic field induced mixing of DX and IX states, characteristic of CQWs.


\section{Theory}



This section presents the theoretical model of exciton spin relaxation in CQWs in the absence of electric bias but in the presence of in-plane magnetic fields. Using  two approaches: semi-phenomenological Lindblad equation approach and the microscopic approach based on Shr\"{o}dinger equation, we will show, that magnetic field dependence of the exciton spin lifetime can by explained by the mixing of DX and IX states.  
Let us consider CQWs with a pair  DX states at energy $E_\mathrm{DX}$ and  
and a pair of IX states at $E_\mathrm{IX}$, Fig. \ref{fig1}~(b, d).
At zero bias  $E_\mathrm{DX}<E_\mathrm{IX}$ due to reduced binding energy of the latter.

The DX state is characterized by a strong electron-hole exchange interaction, while it is vanishingly small for IX state.
The short range component of this interaction $\Delta_0$ splits the bright ($\pm 1$) and the dark ($\pm 2$) exciton momentum states and prevents elastic bright-dark convertion processes.
Spin-flip processes allowed by the selection rules therefore imply simultaneous rotation of both electron and hole spins.
The long-range part of exchange interaction $\Delta$ is wavevector dependent. It is responsible for the Maialle-Andrada e Silva-Sham mechanism of exciton spin relaxation, which dominates in single QWs \cite{MaillaleSham, Vinattieri1994}.
The corresponding spin relaxation time is much shorter than that of IX-bound electron spin relaxation, because exchange interaction is vanishing within IX \cite{Leonard2009, Beian2015, Kowalik, HighPRL2013, Violante}.
We will assume in the following that the spin relaxation rate IXs is governed by its interaction with DXs only.

An in-plane (parallel to the $y$-axis) magnetic field affects the IX-DX coupling via 
magnetic Stark effect \cite{Hopfield}, which may be interpreted via introduction of an effective electric field acting on propagating carriers.
This field is out of CQWs plane and shifts the IX states in full analogy with the real electric field.
The energy scheme and band structure of unbiased CQWs under in-plane magnetic field is shown in Fig. \ref{fig1}~(b, d).
The IX energy dispersion is shifted in the reciprocal space in perpendicular to the magnetic field direction (along the $x$-axis), opposite for the two possible orientations of IX dipole moment \cite{Gorbatsevich, Butov2000}.
The IX and DX parabolic dispersions, characterized by the  effective mass $m_X$ given by the sum of electron and heavy hole in-plane effective masses, intersect at some momentum depending on the DX-IX energy splitting and the shift of the IX dispersion in  magnetic field $k(B)=\pm eBd/\hbar$, Fig. \ref{fig1}~(b). Here $e$ is the electron charge, $d$ is the distance between QWs centres.

Resonant optical pumping generates DXs at the bottom of the dispersion.
However, excitons are accelerated out of the excitation spot by repulsive exciton-exciton interactions.
For DX the characteristic kinetic energy acquired by excitons is given by the interaction energy $6 R_y a_B^2 n$ 
\cite{Ciuti1998}, where $R_y$ and $a_B$ are the exciton Rydberg energy and Bohr radius, respectively, and $n$ is the 2D exciton density, created by the optical pulse.
In realistic conditions this blueshift  is of the same order as the energy difference  between the DX and IX ground states ($\sim 1\,$meV).
This energy is therefore sufficient to reach the range of reciprocal space close to the intersection point of IX and DX dispersions.
Excitons in this range become coupled to the long-living, spin conserving indirect state and define the longest spin relaxation times observed in the experiment. 

To model the experiment we derive the full exciton Hamiltonian, accounting for the electron Zeeman splitting, magnetic Stark effect.
To construct the Hamiltonian we choose a basis of four exciton states, schematically shown in Fig. \ref{fig1}~(b): a pair of spatially direct states and a pair of indirect ones.
Taking into account the exciton spin structure, consisting of four spin states with total momenta projections on the growth axis $\pm 2$ and $\pm 1$, we arrive to the $16\times16$ Hamiltonian matrix form:
\begin{equation}
	\mathrm{H} = \left( \begin{matrix}
		\mathrm{H_{IX,1}} & 0 & \mathrm{J} & 0 \\
		0 & \mathrm{H_{IX,2}} & 0 & \mathrm{J} \\
		\mathrm{J} & 0 & \mathrm{H_{DX,1}} & 0 \\
		0 & \mathrm{J} & 0 & \mathrm{H_{DX,2}}
	\end{matrix} \right)
\end{equation}

Here the spin structure of each excitonic state is given by the diagonal 4x4 blocks $\mathrm{H_{IX(DX),1(2)}}$.
The coupling between them is  given by a single block $\mathrm{J} = J \, \mathrm{I_{4x4}}$, describing spin conserving electron tunneling through the potential barrier between the two QWs.
We neglect the analogous term describing the hole tunneling due to its heavy effective mass in the CQWs growth direction.
This allows  to decouple the two  DX-IX pairs and, without a loss of generality, reduce the problem to the $8\times8$ Hamiltonian:
\begin{equation} \label{Hamiltonian_full}
\mathrm{H} = \left( \begin{matrix}
		\mathrm{H_{IX}} & \mathrm{J} \\
		\mathrm{J} & \mathrm{H_{DX}}
	\end{matrix} \right).
\end{equation}
Motion of dipolar excitons in CQWs is mainly due to repulsion-induced drift rather than pure diffusion \cite{Leonard2009, Rapaport2006}.
Thus, it can be characterised by a quickly fluctuating momentum $\mathbf{q}$ on top of a slowly varying one $\mathbf{K}$, resulting in the total wavevector $\mathbf{k=K+q}$, so that $q \ll K$.
Exciton spin relaxation is induced by the fluctuating part of the long range 
 electron-hole exhange field $\Delta(\mathbf{q}) \propto q$.
We explore two different approaches that allow accounting for this effect.  In the first one we neglect long-range exchange interaction in the DX Hamiltonian, but introduce the fluctuating field via the phenomenological Lindblad superoperator in the Liouville equation for the density matrix. In the second one we directly introduce the fluctuating field in the coherent Hamiltonian part and solve the Shroedinger equation on the timescales shorter than the characteristic momentum scattering time.

In the most general way the DX block in the basis of spin states $(-2, -1, +1, +2)$ reads:
\begin{equation} \label{Hamiltonian_DX}
\mathrm{H_{DX}} = E_\mathrm{DX} (k) \, \mathrm{I_{4x4}} + \left( \begin{matrix}
		-\Delta_0 & \Delta_B & 0 & 0 \\
		\Delta_B & 0 & \Delta(\mathbf{k}) & 0 \\
		0 & \Delta(\mathbf{k})^* & 0 & \Delta_B \\
		0 & 0 & \Delta_B & -\Delta_0
	\end{matrix} \right).
\end{equation}
Here the DX energy $E_{\mathrm{DX}}(k) = {\hbar^2 k^2 / 2 m_X} + E_{\mathrm{DX}}$, $\Delta_0$ is the short range part of the electron-hole exchange interaction,  $\Delta(\mathbf{k})$ is its long range part,  and $\Delta_B = g \mu_B B / 2$. is the electron Zeeman splitting.
 %
Note that we neglect the hole spin splitting due to the small heavy hole $g$-factor.
We also neglect the magnetic Stark effect for DXs as it is linear in electron-hole separation distance.
On the other hand, we keep it in the IX Hamiltonian block:
\begin{align}
	\mathrm{H_{IX}} = E_\mathrm{IX} (k) \, \mathrm{I_{4x4}} 
	 + \left( \begin{matrix}
		0 & \Delta_B & 0 & 0 \\
		\Delta_B & 0 & 0 & 0 \\
		0 & 0 & 0 & \Delta_B \\
		0 & 0 & \Delta_B & 0
	\end{matrix} \right),
\end{align}
where the IX energy $E_\mathrm{IX}(\mathbf{k}) = {\hbar^2 \over 2 m_X} \left( \mathbf{k} - {e \over \hbar} B d \mathbf{e}_x \right)^2 + E_{\mathrm{IX}}$.
The correction to the IX energy appears due to the Landau gauge vector potential $\mathbf{A} = B d \mathbf{e}_x z$, corresponding to the external magnetic field $B \mathbf{e}_y$.
Here $d$ is the mean electron-hole separation within the IX, which may be approximated by the distance between the CQWs centers.
Note that for IXs we neglect both long and short range parts of the electron-hole exchange.

\subsection{Lindblad equation analysis}

In this subsection we use the semi-phenomenological approach based on quantum Liouville equation with the Lindblad superopeoperator in the right-hand part, also called Lindblad equation.
We first assume the linear in $\mathrm{k}$ dependence of the long range exhange $\Delta(\mathrm{k})$ \cite{MaillaleSham}.
It allows us to isolate the slowly varying part of the exchange field, corresponding to the wave vector $K$ from the fast fluctuating part, linear in $q$ : $\Delta(\mathrm{k})=\Delta(\mathrm{K})+\Delta(\mathrm{q})$.
We remove the latter from the coherent Hamiltonian and account for fluctuating field via the Lindblad superoperator term.
The Lindblad equation for the exciton density matrix $\rho$ reads:
\begin{equation} \label{Lindblad}
	{d \rho \over d t} = -{i \over \hbar} \left[ \mathrm{H}, \rho \right] + \mathrm{L}(\rho).
\end{equation}
Here the Lindblad superoperator $\mathrm{L}$ accounts for the DX spin relaxation and decay.
We neglect relaxation and decay processes corresponding to the IX.
The Lindblad term $\mathrm{L}(\rho)$ also may be represented in a block matrix form:
\begin{equation}
\mathrm{L}(\rho) = \left( \begin{matrix}
0 & -\rho_{\mathrm{DX-IX}} \gamma_\mathrm{DX}/2 \\
-\rho_{\mathrm{IX-DX}} \gamma_\mathrm{DX}/2 & \mathrm{L_{DX}}(\rho_\mathrm{DX-DX})
\end{matrix} \right).
\end{equation}
Here $\rho_{\mathrm{DX-IX}}$ and $\rho_{\mathrm{IX-DX}}$ are the top-right and bottom-left 4x4 blocks of the total density matrix, which decay at twice slower rate than the DX density $\gamma_\mathrm{DX}/2$.
The $\mathrm{L_{DX}}$ block, which describes the fast relaxation of the direct part, reads:
\begin{widetext}
\begin{equation}
\mathrm{L_{DX}}(\rho_\mathrm{DX-DX}) = - {\rho_\mathrm{DX-DX} \gamma_\mathrm{DX}} + \left( \begin{matrix}
0 & 0 & 0 & 0 \\
0 & \left(\rho_{\mathrm{DX-DX}}^{+1,+1}-\rho_{\mathrm{DX-DX}}^{-1,-1}\right)\gamma_{ex} & - \rho_{\mathrm{DX-DX}}^{-1,+1}\gamma_{ex} & 0 \\
0 & - \rho_{\mathrm{DX-DX}}^{+1,-1}\gamma_{ex} & \left(\rho_{\mathrm{DX-DX}}^{-1,-1}-\rho_{\mathrm{DX-DX}}^{-1,-1}\right)\gamma_{ex} & 0 \\
0 & 0 & 0 & 0
\end{matrix} \right),
\end{equation}
\end{widetext}
where $\gamma_{ex} = (2\tau_{ex})^{-1}$ is the rate of spin relaxation due to the Maialle-Andrada e Silva-Sham mechanism.
We assume that the hole spin relaxation is by far the fastest process in the system and take as initial condition for the Lindblad evolution equation the density matrix with only two non-zero elements $\rho_\mathrm{DX-DX}^{+1,+1}=\rho_\mathrm{DX-DX}^{-2,-2} = n/2$, which describes the excitonic system of density $n$, pumped with a $\sigma+$ polarized optical pulse, with a fully relaxed heavy hole spin.

The solution of equation (\ref{Lindblad}) yields the dynamics of the Kerr rotation angle, measured in the experiment.
The effect itself is produced by spin-dependent exciton resonance shifts, stemming from exciton-exciton exchange interactions.
Both direct and indirect components contribute to the value of Kerr rotation angle.
The value of the Kerr rotation angle $\delta \theta$ is a sum of two contributions, linear in bright DX and bright IX spin polarizations $\rho_\mathrm{DX-DX}^{+1,+1}-\rho_\mathrm{DX-DX}^{-1,-1}$ and $\rho_\mathrm{IX-IX}^{+1,+1}-\rho_\mathrm{IX-IX}^{-1,-1}$ respectively \cite{Nalitov2013}.
Furthermore, the coefficients before the two polarizations, given by the Coulomb carrier exchange, weakly depend on the electron-hole separation distance and may be assumed equal for IX and DX contributions, allowing us to write:
\begin{equation} \label{Kerr}
\delta \theta \sim \rho_\mathrm{DX-DX}^{+1,+1}-\rho_\mathrm{DX-DX}^{-1,-1}+\rho_\mathrm{IX-IX}^{+1,+1}-\rho_\mathrm{IX-IX}^{-1,-1}.
\end{equation}

Kerr rotation angle decay, as well as the relative spin polarisation, may be extracted from the solution of Eqs. (\ref{Lindblad}, \ref{Kerr}) as functions of the magnetic field $B$.
We fit the experimental data
assuming the exciton effective mass $m_X = 0.22 m_0$ \cite{Lozovik2002} and the interwell distance $d=12\,$nm.
The electron Lande factor $g = 0.1$ is obtained from experiments, and the bright-dark splitting is $\Delta_\mathrm{0}=70$~$\mu$eV \cite{Blackwood1994}.
In the simulation we neglected the coherent electron-hole exchange field $\Delta(\mathrm{K})$ compared with other fields in the system for both DX and IX.
The decay time obtained with the  fit parameters  $\Delta E = E_\mathrm{DX} - E_\mathrm{IX}=1.5$~meV, $J=0.15$~meV,  $K=90$~$\mu$m$^{-1}$,  $1/\gamma_{DX}=2$~ns,   $1/\gamma_{ex}=400$~ps, is shown in Fig. \ref{figVg0}~(c) by solid line.  
It has a maximum in the vicinity of $B=5\,$T, which corresponds to the maximum DX-IX mixing.
One can see, however, that the relation between the decay times at the peak and at $B=0$ is limited by $2$.
Indeed,  in the ideal case of negligible DX-IX coupling at $B=0$ where the exciton spin relaxation is given by that of DXs, the longest possible decay time is achieved whilst the DX and IX modes are resonant.
Excitons in this case are half-indirect and thus lose the spin polarization at the twice reduced rate, according to the solution of the Lindblad equation.
In the realistic case, where the coupling $J \neq 0$, the exciton spin dynamics is affected by the IX admixture even at $B=0$, therefore this approach yields $2$ as the upper limit for the relation 
$\tau_s(B=5)/\tau_s(B=0)$, whereas the measured value is close to $3$.

The main reason for the discrepancy between this result and the experimental data lies in the phenomenological nature of $\gamma_{ex}$, introduced as the relaxation rate of the direct exciton part.
Note that this parameter differs from the exciton spin depolarization time in a single QW of the same width as those composing CQWs.
The classical Dyakonov-Perel picture gives an insight to this difference.
As long as the characteristic exciton transport time $\tau$ is longer than the electron tunneling time $(J/\hbar)^{-1}$, the exciton loses its spin as a whole, rotating in a stochastic effective magnetic field between scattering events, rather than losing it via its DX and IX components independently.
Our further microscopic analysis of the spin relaxation gives similar qualitative result as the semi-phenomenological Lindblad equation-based model, but allows to improve the quantitative agreement between the theory and the experiment.

\subsection{Microscopic analysis}

Assuming a generic spin relaxation mechanism stemming from the spin precession in a stochastically fluctuating field, the spin relaxation rate $\gamma_s=1/\tau_s$ scales with both characteristic value of the fluctuating field $\Omega$ and the fluctuation time scale, given by the momentum relaxation rate $\tau$ via Dyakonov-Perel formula  $\gamma_s = \Omega(B)^2 \tau$ \cite{Dyakonov1971,MaillaleSham}.
In principle, both parameters depend on the degree of DX-IX coupling: $\Omega$ scales with the electron-hole overlap, while $\tau$ in drift-diffusion regime depends on the value of the exciton dipole moment.
Here we focus on the variation of $\Omega$ with magnetic field $B$, which allows us to explain the experimental measurements.
Instead of the Lindblad equation we solve the Shr\"{o}dinger equation, taking the full Hamiltonian (\ref{Hamiltonian_full}) with the field, stemming from the long-range electron-hole exchange $\Delta(\mathrm{k})$.
This approach is valid on  the timescales shorter than the momentum relaxation time, as the effective field is momentum dependent and its absolute value is in the first approximation linear in the momentum value \cite{BirPikus}.


Taking a wavefunction, describing a DX with the fully relaxed hole spin and the electron spin $s_e=-1/2$, as initial condition, one may trace the corresponding Kerr rotation angle $\delta \theta (t)$.
Treating it as bright exciton subspace pseudospin projection, we numerically extract the characteristic frequencies of its rotation $\Omega(B)$ from the Fourier transform of $\delta \theta (t)$. 
Assuming $\Omega \tau \ll 1$, we estimate the exciton spin decay rate as $\gamma_s = \Omega(B)^2 \tau$. The parameters of the numerical calculation are the same as for the Lindblad equation.

This approach yields up to $4$ times increase of the measured spin relaxation time 
$\tau_s=1/\gamma_s$ at the resonance of DX and IX energies, as the stochastic rotation frequency can be twice 
lower for a coupled DX-IX state in comparison with a pure DX state.
The fit of the experimental data using this microscopic approach (dashed line in Fig. \ref{figVg0} (c)) is therefore more accurate in comparison with the semi-phenomenological approach based on the Lindblad equation (solid line), 
even though it does not take into account the dependence  of the transport time $\tau$ on the magnetic field.

\section{Conclusions}

We have studied spin dynamics of excitons in CQWs in the presence of crossed magnetic and electric fields using time-resolved Kerr rotation spectroscopy. 
Two qualitatively different regimes of spin decoherence are identified, depending on the strength of the electric field, applied along the growth axis.
In the presence of the gate voltage, such that IX becomes  the lowest energy  exciton state of the system, the inhomogeneous spin coherence time is found to be inversely proportional to the magnitude of the in-plane magnetic field. 
This behaviour is understood in terms of the inhomogeneous distribution of g-factors, typical for 
QW structures.
Inevitably, such distribution leads to the broadening of the spin precession frequency distribution between excitons, and thus linear dependence of the spin dephasing rate on the magnetic field. 
This inhomogeneity seems to be stronger for excitons than for electrons, probably due to stronger localisation of excitons, which are heavier particles.
Completely different mechanisms dominate the spin coherence in symmetric CQWs, when zero, or small  electric field is applied, so that  IX energy is higher than that of DX state.
In this regime, we have found manifestations of the quantum confined magnetic 
Stark effect in the exciton spin relaxation time dependence on the in-plane magnetic field in CQWs.
The strongly non-monotonous behaviour of the spin lifetime that may seem counter-intuitive finds its explanation in the magnetic field induced mixing of DXs and IXs due to the shift of the IXs dispersion curve.
This is the signature of the  magnetic Stark effect.
The magnetic Stark effect appears to be a convenient tool of exciton spin engineering, 
that may complement traditional quantum confined Stark effect in the structures where inhomogeneity 
is important.

\emph{Acknowledgments.} We are grateful to K.~V.~Kavokin, A.~P.~Dmitriev, N.~A.~Gippius and M.~I.~Dyakonov for
valuable discussions and acknowledge the support of EU ITN INDEX
PITN-GA-2011-289968. LVB  was supported by DOE Award
DE-FG02-07ER46449. AK
acknowledges the support from the Russian Ministry of Education and
Science, Contract No 11.G34.31.0067. MN acknowledges the support
from the Polish National Science Center under decision
DEC-2013/09/B/ST3/02603.

\bibliography{IX}

\end {document}